\documentclass[twocolumn,aps,preprintnumbers,amsmath,amssymb,superscriptaddress,nofootinbib]{revtex4}

\usepackage{slashed}
\usepackage{epsfig,latexsym,cancel,amssymb,amsmath,verbatim,mathrsfs}
\usepackage{color}
\usepackage{graphicx,subfigure,multirow}

\usepackage{dcolumn}
\usepackage{bm}
\usepackage[all]{xy}
\usepackage{color}
\usepackage[usenames,dvipsnames]{xcolor}
\usepackage[unicode=true,pdfusetitle,
 bookmarks=true,bookmarksnumbered=false,bookmarksopen=false,citecolor=Turquoise,
 breaklinks=false,pdfborder={0 0 1},backref=false,colorlinks=true,pdfpagemode=FullScreen]
 {hyperref}
\usepackage{amssymb}
 \usepackage{longtable}
\usepackage{epsfig}
\usepackage{amsmath}
\usepackage{pstricks}
\usepackage{bbold}
\usepackage{slashed}
\usepackage{amssymb}
\usepackage{graphicx}
\usepackage{hyperref}
\usepackage{verbatim}
\usepackage{multirow}
\usepackage{chemarrow}
\usepackage{hhline}
\makeatletter

\newcommand{\beq}{\begin{equation}}
\newcommand{\eeq}{\end{equation}}
\newcommand{\bea}{\begin{eqnarray}}
\newcommand{\eea}{\end{eqnarray}}
\newcommand{\GeV}{\,\text{GeV}}
\newcommand{\TeV}{\,\text{TeV}}

\newcommand{\ET}{\mbox{$\not \hspace{-0.10cm} E_T$}}
\newcommand{\VET}{\mbox{$\not \hspace{-0.10cm} \vec{E}_T$}}

\begin{document}

\preprint{
\begin{minipage}[b]{1\linewidth}
\begin{flushright}
APCTP-Pre2015-021\\
IPMU15-0120\\
CTPU-16-09
 \end{flushright}
\end{minipage}
}

\title{ Revealing the jet substructure in a compressed spectrum}

\author{Chengcheng Han}
\email[]{hancheng@apctp.org}
\affiliation{Asia Pacific Center for Theoretical Physics, Pohang, 790-784, Korea}
\affiliation{Kavli IPMU (WPI), The University of Tokyo, Kashiwa, Chiba 277-8583, Japan}
\author{Myeonghun Park}
\email[]{parc.ctpu@gmail.com}
\affiliation{Asia Pacific Center for Theoretical Physics, Pohang, 790-784, Korea}
\affiliation{Kavli IPMU (WPI), The University of Tokyo, Kashiwa, Chiba 277-8583, Japan}
\affiliation{Center for Theoretical Physics of the Universe, Institute for Basic Science (IBS), Daejeon, 34051, Korea}

\date{March 28, 2016}

\begin{abstract}
The physics beyond the Standard Model with parameters of the compressed spectrum is well motivated both in a theory side and with phenomenological reasons, especially related to dark matter phenomenology.
In this letter, we propose a method to tag soft final state particles from a decaying process of a new particle in this parameter space.
By taking a supersymmetric gluino search as an example, we demonstrate how the Large Hadron Collider experimental collaborations can improve a sensitivity in these non-trivial search regions.

\end{abstract}

\maketitle
{\it Introduction.}---After discovering the Higgs boson at the Large Hadron Collider (LHC), the most important question is whether the scale of the physics beyond the Standard Model (BSM) is within the coverage of the LHC.
As current LHC searches have been pushing away a possible energy scale of the BSM,
most part of parameter space around  $\mathcal{O}(1)\TeV$  mass scale in various models of BSM have been ruled out.
But so far existing LHC analyses lose sensitivities in the compressed spectrum region where the difference of masses among new particles are negligible compared to their mass scale.
Major difficulties in those analyses are from tagging soft particles over Standard Model Quantum Chromodynamics (QCD) backgrounds. 
The parameter space of compresses spectrum is common in various new physics scenarios.
In a supersymmetric framework (SUSY), the compressed spectrum is naturally predicted by various SUSY  breaking models \cite{Murayama:2012jh, Nakayama:2013uta}. For the extra dimension models like the Universal extra dimensional one (UED),
 the degenerate spectrum in Kaluza Klein (KK) modes is usually predicted and only radiative correction can give limited  mass difference between different KK particles\,\cite{Cheng:2002iz}.
From the phenomenological point of view, BSM with the compressed spectrum is favored.
Dark matter annihilate processes in this parameter space case can easily satisfy a relic density compatible to current observations through the co-annihilation process as in the bino-wino or bino-gluino co-annihilation region \cite{well-temper,Profumo:2004wk}.
Due to the importance of compressed spectrum, there have been lots of studies\,\cite{compress-search}. 
Unlike previously suggested analyses, we propose a method to tag soft jets from decays of BSM particles.  

{\it A fat-jet for compressed spectrum.}---
The event topology of our consideration is a three-body decay channel where a BSM particle $A$ decays into two quarks and another BSM particle $B$,
\beq
A \to q+q' + B \, ,
\eeq
with a case of compressed mass spectrum,
\beq
\Delta m \equiv m_A - m_B \ll m_A .
\eeq
When the mass splitting is negligible compared to the mass of $A$, most of the energy from this decay flows into $B$ making each quark $q,q'$ too soft to be tagged as an isolated jet. This type of a decay is very common in BSM, some examples in SUSY with a gluino decaying into a neutralino or a heavier chargino/neutralino decaying into the lighter neutralino when a mass splitting is less than $W/Z/H$ bosons.
We should note that current collider searches do not provide satisfactory results in this case, especially in search channels of multi-jet\,\cite{Aad:2014wea} and of mono-jet+missing transverse energy ($\ET$) channel \cite{monojet} that barely provides limits on neutralino mass around $600\GeV$.
\begin{figure*}[t!]
\includegraphics[width=0.4\textwidth]{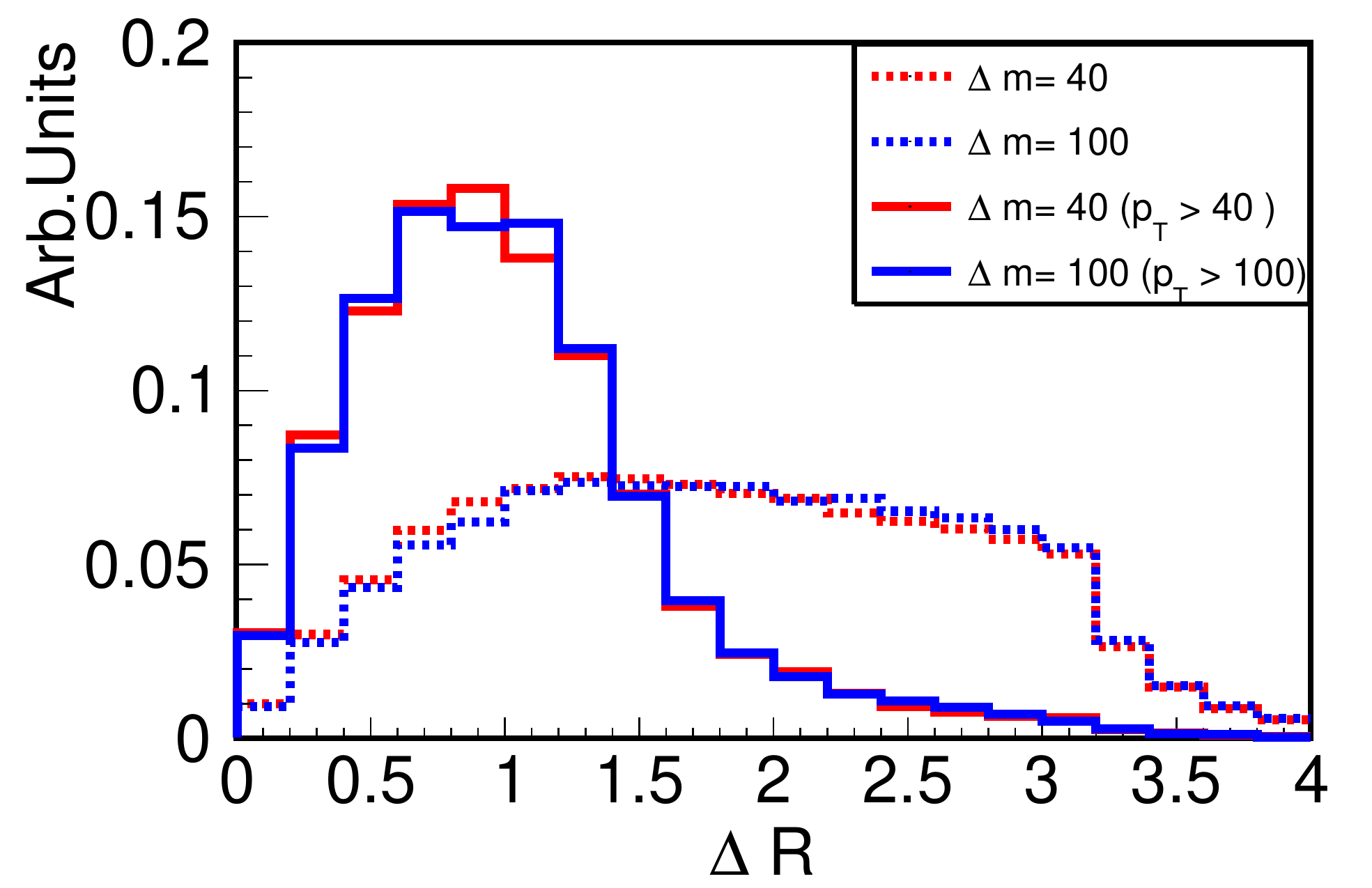} \qquad
\includegraphics[width=0.4\textwidth]{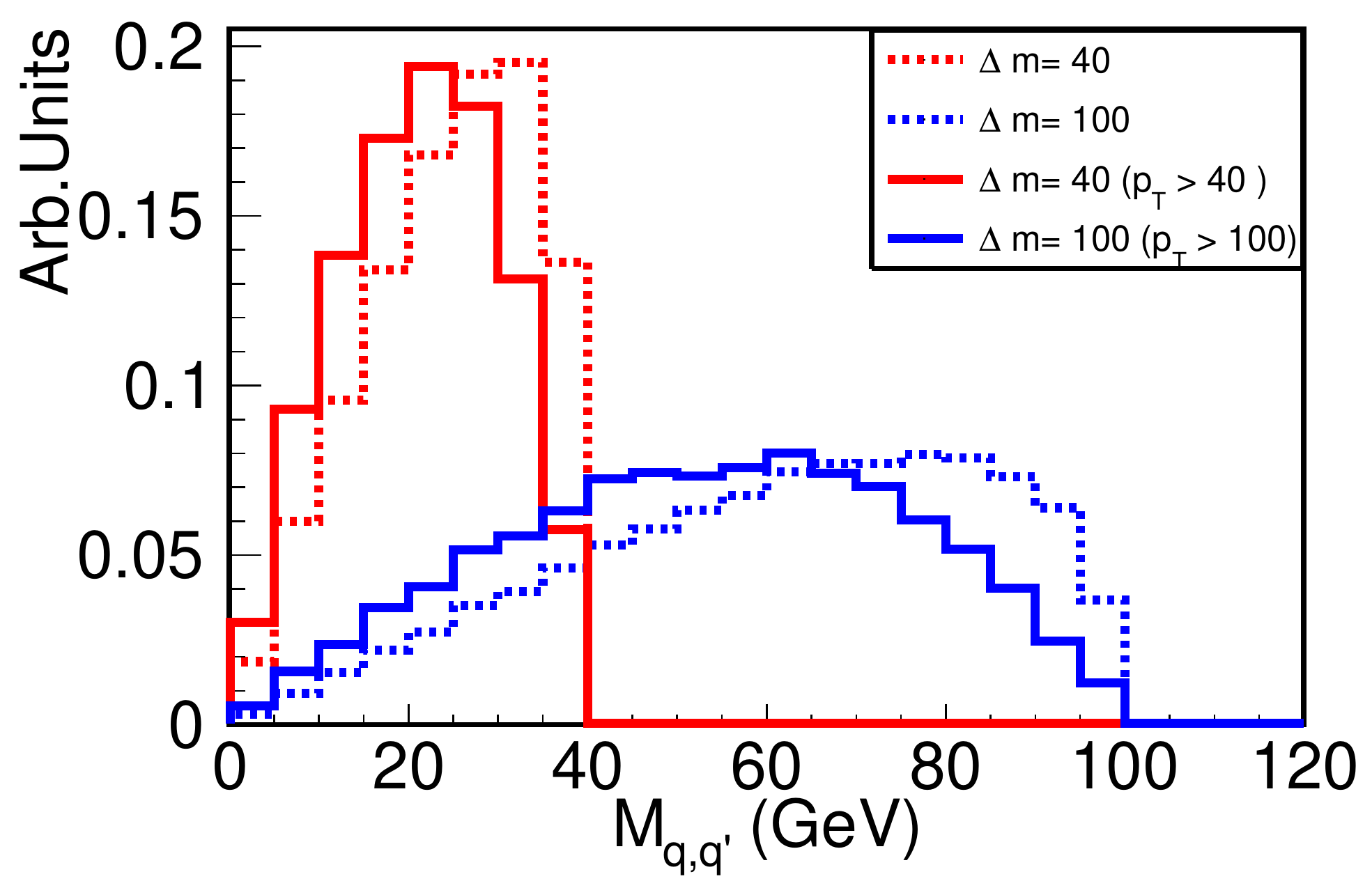}
\centering
\caption{\label{parton}We show distributions of a distance $\Delta R$ between two quarks from a gluino decay (a) and an invariant mass $M_{qq'}$ (b) using a parton level Monte Carlo (MC) simulation. Here we choose $\Delta m=40, 100 \GeV$, solid lines with a requirement of $p_{T(\textrm{FJ})}> \Delta m$ and dotted lines without $p_{T(\textrm{FJ})}$ requirement.}
\end{figure*}
Thus we introduce a method to tag jets from $A$ decay with suppressing the Standard Model backgrounds.
To identify quarks in the above process, we propose a ``fat-jet (FJ)" to capture quarks $(q,q')$ as a single clustered object. The cone size of a fat-jet will be determined purely by the kinematics of a decaying process at the leading order QCD. A distance $\Delta R$ between two quarks would be,
\beq
\Delta R\approx \frac{1}{\sqrt{z(1-z)}} \frac{M_{qq^\prime}}{p_{T(\textrm{FJ})}}\, ,
\label{eq:deltaR}
\eeq
where $p_{T(\textrm{FJ})}$ is the total transverse momentum of $(q,q')$ (a fat-jet),  $z$ is a $p_T$ fraction of $A$ that $q$ obtains.
In a case of a three body decay, due to a symmetric feature of $(q \leftrightarrow q')$, $z \simeq 1/2$. $M_{qq'}$ denotes an invariant mass of $q,q'$.
With limited data, $M_{qq'}$ will be localised around the peak $P_{qq'}$ of an invariant mass distribution \cite{Cho:2012er},
\bea
P_{qq'} &&= \left[\frac{m_A^2+m_B^2}{3} \left(2-\sqrt{1+\frac{12 m_A^2 m_B^2}{(m_A^2+m_B^2)^2}}\right)\right]^\frac{1}{2} \nonumber \\
            &&\underset{\Delta m \ll m_A}{\longrightarrow} \frac{\Delta m}{\sqrt{2}},
\eea
To select signal events over backgrounds, we require a cut on a transverse momentum $p_T$ of a fat-jet.
With $p_{T(\textrm{FJ})}>\Delta m$, we have $\Delta R\lesssim \sqrt{2}$ from eq.\,(\ref{eq:deltaR}).
Thus we set a radius R of a fat-jet to $1.5$ to capture two quarks from $A$ decay. In FIG.\,(\ref{parton}), we show a distance between two quarks from $A$ decay and their invariant mass using a MC simulation at a parton level. We generate SUSY gluino pair productions and their decays into two quarks and a neutralino. The neutralino mass is fixed to $600\GeV$ and gluino mass varies according to the mass difference $\Delta m$.  As we see, we can capture most of two quarks from a gluino decay with a fat-jet radius $R=1.5$ after a cut on a fat-jet, $p_{T(\textrm{FJ})}>\Delta m$.
Our approach has a benefit of statistical gain in a requirement on $p_T$ of an initial state radiation (ISR) over mono-jet+$\ET$ searches.
We only require ISR $p_T$ as large enough for $A$ to be boosted resulting in $p_{T(\textrm{FJ})}> \Delta m$.

\begin{figure*}[t!]
 \includegraphics[width=0.4\textwidth]{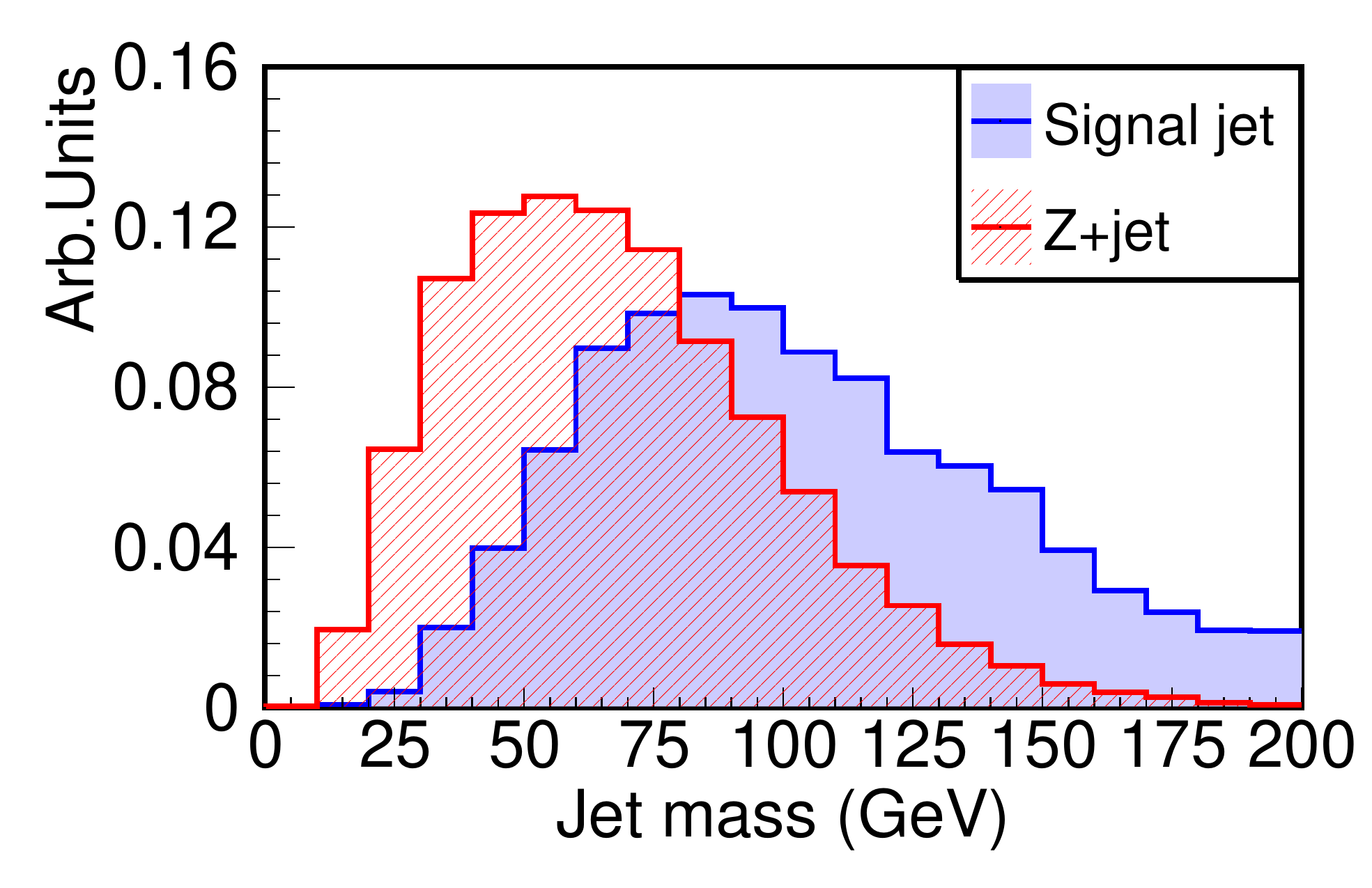}\qquad
 \includegraphics[width=0.4\textwidth]{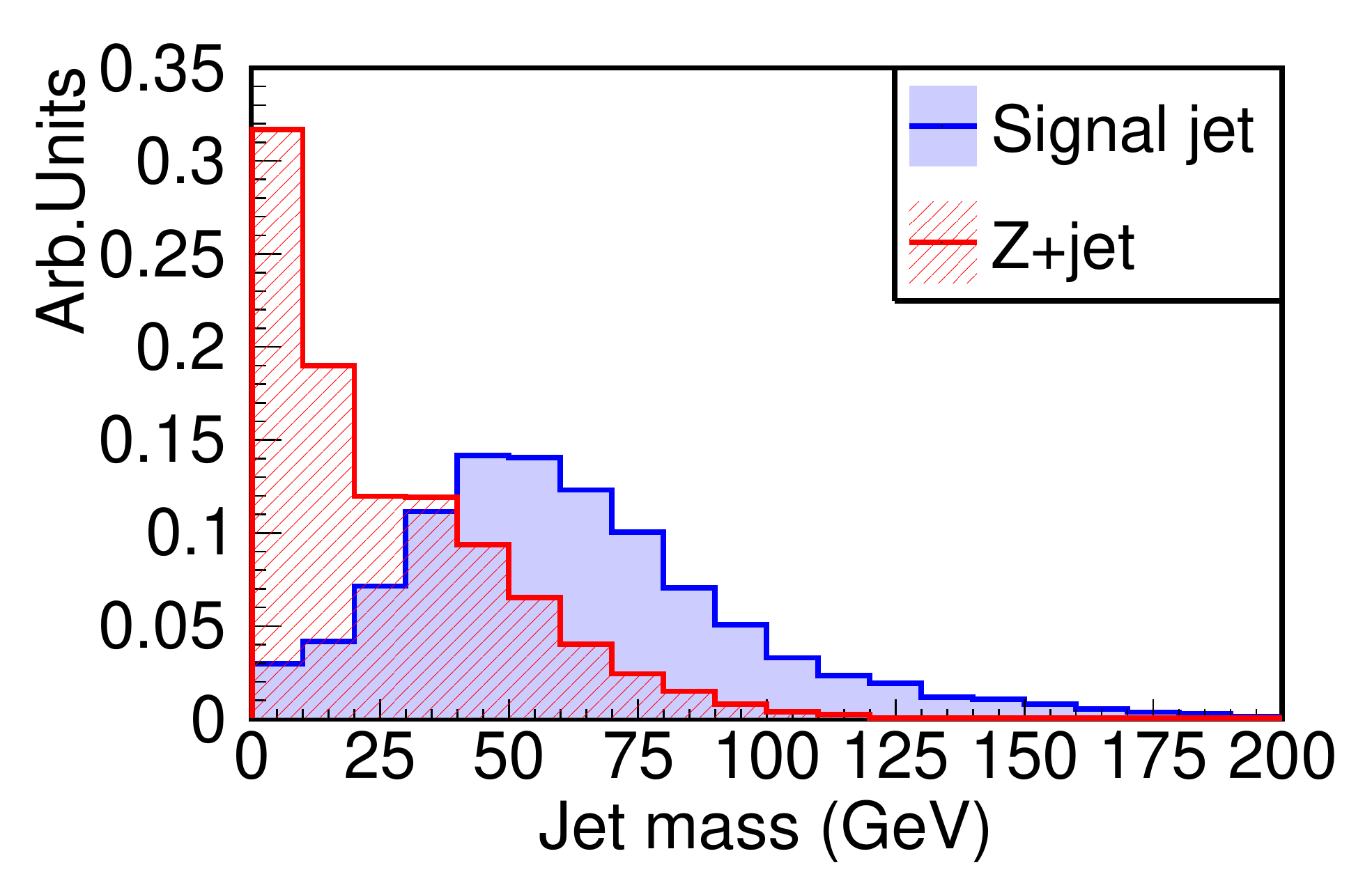}
\caption{We plot fat-jet mass distributions of signal fat-jet and background QCD fat-jet before MDT (a) and after MDT (b) at a reconstructed level. For a signal process we use $\Delta m=m_{\tilde g}-m_{\tilde \chi_1^0} =100\GeV$ with $m_{\tilde \chi_1^0}=600\GeV$.   }
\label{fig:before}
\end{figure*}

With introducing a large size jet, we have issues from soft QCD corruptions.
A normal QCD jet (NLO) is approximately to
\beq
\sqrt{<M_J^2>_{\textrm{NLO}}}\, \approx 0.2 p_J R\, ,
\eeq
with a numerical factor 0.2 including colour charges \cite{Ellis:2007ib}. For an example with a mass splitting of $\Delta m = 100\GeV$, a normal QCD jet attains a mass of $\sqrt{<M_J^2>} \sim 30\GeV$ with $p_{T(\textrm{FJ})}> \Delta m$. But as we show in FIG.\,(\ref{fig:before}), due to underlying events, background jets become more massive.
To remove these additional contributions on a jet mass and to sort out background QCD fat-jet\footnote{We use a term QCD fat-jet for the background QCD fat-jet.}, we utilize a grooming technique and a jet substructure variable.

{\it Signal fat-jet v.s.\,QCD fat-jets.}---We note that a signal fat-jet is a two-prong jet while QCD fat-jet gets most of its energy from a single prong and obtains a volume by soft QCD radiations.
A lot of studies have been performed to distinguish two-prong jets from the normal jets \cite{MDT,N-sub,kt-split} with focusing on large $p_T$ regions where preferences come from the motivation of boosted objects.
Here we show how one can adapt those techniques to the compressed spectrum in the region of moderate $p_T$.
To show the difference between signal fat-jets and QCD fat-jets, we simulate supersymmetry gluino pair productions as in the previous section and Z+jet productions for backgrounds. Both signals and backgrounds are generated by Madgraph\_aMC@NLO\,\cite{mad} with PYTHIA 6.4.28\,\cite{pythia}.
We apply ATLAS AUET2B\,\cite{AUET2B,Aad:2012fza} tune with PDF set CT6L\cite{cteq}. For a detector simulation, we use Delphes\,v3\,\cite{delphes} interfaced with FastJet\,v3.0.6\,\cite{fastjet} for a jet clustering with ATLAS detector parameters.
Fat-jets are clustered according to Cambridge-Aachen(C/A) algorithm\,\cite{CA} with $R=1.5$.

To decontaminate a fat-jet from soft QCD corruptions, we implement the Mass Drop Tagger (MDT)\,\cite{MDT} out of various grooming techniques\,\cite{MDT, trimming, pruning}.
We choose a parameters for MDT same as in BDRS Higgs tagger \cite{MDT} since our spectrums $\Delta m$ are about the similar order of the Higgs' mass. In FIG.\,(\ref{fig:before}) we show jet mass distributions of signal fat-jets and QCD fat-jets before and after passing MDT. There is a huge overlapping in jet mass distributions between signal fat-jets and QCD fat-jets before MDT.  However, after MDT procedure, the signal jet clearly shows a peak around 60 GeV, and the distribution is similar to our parton level analysis in FIG.\,(\ref{parton}) while QCD fat-jet becomes lighter.
In short, our result shows that MDT method effectively removes soft-QCD corruptions in ``non-boosted" $p_T$ region.
 In addition to this different shape in jet mass distributions between the signal and the background, QCD fat-jets easily fail the MDT procedure since QCD fat-jet is characterized primarily by a single dense core of energy deposits surrounded by soft radiations that is not compatible to symmetric conditions. Thus MDT itself can act as a good analysis cut.

\begin{figure}[t!]
\centering
\includegraphics[width=0.4\textwidth]{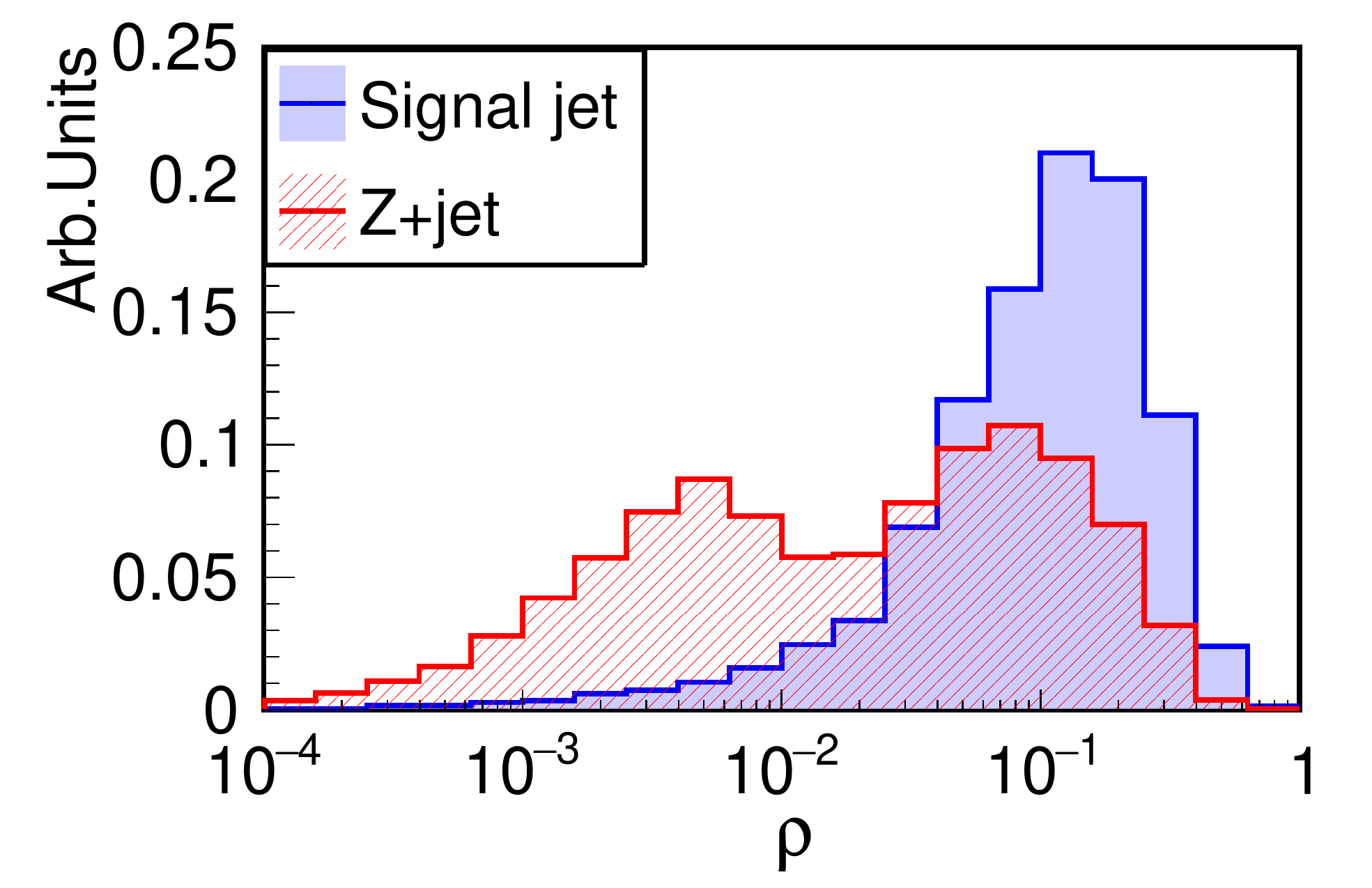}
\caption{\label{rho}$\rho$ distributions after a MDT procedure at a reconstruction level. The study point of a signal process is the same as FIG.(\ref{fig:before}).}
\end{figure}

To increase a tagging efficiency, we use a dimensionless parameter $\rho=m^2_j/\left(p^2_{Tj}R^2\right)$ to impose a cut on fat-jets. This $\rho$ parameter characterizes a splitting of two quarks inside a fat-jet.
For a signal fat-jet with a radius $R$, $\rho$ parameter becomes
\beq
\rho\sim z(1-z) \frac{R^2_{qq'}}{R^2},
\eeq
from eq.(\ref{eq:deltaR}). $R_{qq'}$ is the distance of the two quarks. For signal fat-jets, we can easily estimate a theoretical value of $\rho\sim 0.2$ with our choice of $p_{T(\textrm{FJ})}$ and $R$ in the previous section. For QCD fat-jets, we take a quark-initiated jet as an example since most of jets in major background of Z+jet are quark-initiated. The leading order distribution of $\rho$ from a jet consisting of two quarks\,\cite{jetmass}:
\beq
\frac{\rho}{\sigma} \frac{ d\sigma}{ d \rho}=\frac{\alpha_s C_F}{\pi}\Big[\Theta(\rho-y_*)\ln\frac{1}{\rho}
 +\Theta(y_*-\rho)\ln\frac{1}{y_*}-\frac{3}{4}\Big]\,.
\eeq
The QCD fat-jet has a soft large splitting or hard small splitting which provides a small $\rho$ value.
In FIG.\,(\ref{rho}), a distribution of signal fat-jets clearly has a peak around 0.2 as our estimation. The population of $\rho$ in QCD fat-jet locates less than 0.1. This encourages us to impose a cut  $\rho>0.1$.
The cut on $\rho$ is independent on $p_{T(\textrm{FJ})}$ and provides an orthogonal constraint to the cut on fat-jet mass which is proportional to $p_{T(\textrm{FJ})}$.
 In Tab.\,(\ref{tab1}) we show a tagging efficiency ($\epsilon_{\textrm{tag}}$) and a fake efficiency ($\epsilon_{\textrm{fake}}$) of tagging fat-jet. 
Signal fat-jet indeed has a much higher tagging efficiency compared to QCD fat-jet. 
\begin{table}[th]
\begingroup
\renewcommand{\arraystretch}{1.2} 
\begin{tabular}{|c||c|c|c|c|c|c|c|c|}
\hline
  BP & \multirow{2}{*}{cuts}       & Signal &  $Z+j$ \\
$m_{\tilde \chi_1^0}=600$                  &                        & ($\epsilon_{\textrm{tag}}$)  & ($\epsilon_{\textrm{fake}}$) \\[1mm]
\hline
BP1 & $10< m_\textrm{FJ}<40$        & 0.75      & 0.55    \\[1mm]
    $\Delta m=40$          & Additional $\rho >0.1$ cut & 0.40      & 0.18   \\[1mm]

\hline
BP2 & $40< m_\textrm{FJ}<100$        & 0.62      & 0.22    \\[1mm]
$\Delta m=100$     & Additional $\rho >0.1$ cut  & 0.42      & 0.12    \\
\hline
\end{tabular}
\endgroup
\caption{\label{tab1}Tagging/faking efficiencies on a fat-jet. We have two benchmark points (BP1, BP2) depending on a mass splitting.
Mass unit is a$\GeV$.}
\end{table}

{\it Improvement on sensitivities.}---We show a benefit from our method with a fat-jet by applying it to existing ATLAS gluino search of the LHC 8TeV Run 1\,\cite{Aad:2014wea}.
We select a signal region (2jm) in ATLAS search which is the most sensitive region in our benchmark points.
An analysis on (2jm) requires at least two hard jets in final states and cuts are shown in Tab.\,(\ref{tab2}).

 For our signal, we generate gluino pair production up to one jet matching using MLM scheme\,\cite{mlm} to be consist with ATLAS analysis\,\cite{Aad:2014wea}.
 For backgrounds, we simulate three major backgrounds: $Z(\nu\bar\nu)+$jets, $W(\ell\nu)+$jets and $t\bar{t}+$jets.
 For the $Z(\nu\bar\nu)+$jets, $W(\ell\nu)+$jets events, we generate Z/W production adding from 1 to 3 jets.
 For the $t\bar{t}+$jets production, we generate $t\bar{t}$ with up to 2 jets. For the hard QCD backgrounds, because of the requirement of the large missing energy and  $m_{\textrm{eff}}$, those QCD backgrounds are reduced to be less than $0.1\%$ of the total backgrounds \cite{Aad:2014wea}, so we can safely disregard its contribution to the total backgrounds.  Additional jet in background events is considered through MLM matching procedure \cite{Mangano:2006rw}.
\begin{table}[th]
\begingroup
\renewcommand{\arraystretch}{1.3} 
\begin{tabular}{|c|c|}
\hline  Signal region (2jm)     &  cuts   \\
\hline  $\ET$       & $>160\GeV$             \\
\hline  $p_T(j_1)$         & $> 130\GeV$           \\
\hline  $p_T(j_2)$          & $>60\GeV$            \\
\hline  $\min\Delta \phi(p_{T(j_{1,2,(3)})},\VET)$          & $> 0.4$            \\
\hline  $\ET/\sqrt{H_T}$         & $> 15\GeV^{1/2}$            \\
\hline  $m_{\textrm{eff}}$     & $> 1200\GeV$            \\
\hline
\end{tabular}
\endgroup
\caption{\label{tab2}ATLAS cut flows for signal region (2jm).}
\end{table}
Before applying ATLAS analysis, we require a fat-jet analysis with clustering particles according to C/A of R=1.5 and select a fat-jet with following criteria:
\begin{itemize}
\item[(1)] Apply a cut $p_{T(\textrm{FJ})} > \Delta m$ on fat-jets (after MDT). 
\item[(2)] Choose the fat-jet with the largest $\rho$ and mark this fat-jet as a candidate for a signal fat-jet.
\end{itemize}
To show the effectiveness of above criteria, we define following two efficiencies:
\begin{eqnarray}
\epsilon_1=\frac{N_{\textrm{signal}}}{N_{\textrm{total}}},\quad \epsilon_2=\frac{N_{\textrm{candidate}}}{N_{\textrm{signal}}}\, ,
\end{eqnarray}
where $N_\textrm{total}$ is the number of generated signal events and $N_\textrm{signal}$ denotes a number of events that contains a fat-jet from a gluino decay in a reconstructed fat-jet list after MDT.
The $N_\textrm{candidate}$ is the number of events where a fat-jet from a gluino decay is taken as the candidate through above criteria.
Thus $\epsilon_1$ gives the fraction of events containing the signal fat-jets and $\epsilon_2$ shows the efficiency of finding signal fat-jet.
We find nearly half of the signal events contains the signal fat-jet and our method can reconstruct $60\%-70\%$ signal fat-jets as in Tab.(\ref{tab3}).
\begin{table}[th]
\begingroup
\renewcommand{\arraystretch}{1.3} 
\begin{tabular}{|c|c|c|}
\hline           & $\Delta M=40\GeV$    &  $\Delta M= 100\GeV$  \\
\hline    $\epsilon_1$         & 0.64      & 0.56    \\
\hline  $\epsilon_2$        & 0.58  & 0.74\\
\hline
\end{tabular}
\endgroup
\caption{\label{tab3}Selection efficiency for the signal events.}
\end{table}
\begin{table}[!h]
\begingroup
\renewcommand{\arraystretch}{1.3} 
\begin{tabular}{|c|c|c|c|c|c|c||c|c||c|c|}
\hline
\multicolumn{ 3}{|c|}{Methods}   & Z+js   & W+js   & $t\bar{t}$+js & Total & BP1 & $\sigma$ & BP2 & $\sigma$ \\
\hline
\multicolumn{ 3}{|c|}{ATLAS}   & 430      & 216   &  47  &  693  & 74 & 1.34 & 57 & 1.03 \\
\hline
With & \multirow{2}{*}{$\Delta m$} & $40$ & 48 & 31 & 7 & 86 & 27 & 2.44 &\multicolumn{ 2}{c|}{$-$} \\
\cline{3-11}
          FJ                            &  & $100$ & 21 & 18 & 11 & 50 &\multicolumn{ 2}{c||}{$-$}   & 17 & 2.15\\
\hline
\end{tabular}
\endgroup
\caption{\label{tab5} Expected number of events and corresponding significance of the LHC $8\TeV$ with a luminosity of $20.3\textrm{fb}^{-1}$.
We refer readers to Tab.\,(\ref{tab2}) for cuts of (2jm). The signal cross section for  $\Delta m=40$ GeV is  0.83pb and for $\Delta m=100$ GeV is  0.43pb.}
\end{table}
After preselecting events with mass window cut and $\rho$ cut on a fat-jet as in Tab.\,(\ref{tab1}), we recluster particles with anti-$k_t$\,\cite{anti-kt} of $R=0.4$ according to the ATLAS analysis and apply cuts in Tab.\,(\ref{tab2}).
We compare the performance of our procedure of a fat-jet analysis with ATLAS analysis in Tab.\,(\ref{tab5})\footnote{The backgrounds without fat jet tagging are scaled to the 
background events reported by \cite{Aad:2014wea}}.
The systematic uncertainty here is assumed to be $7\%$  from in ATLAS (2jm) search\,\cite{Aad:2014wea},
We would like to stress a point that we do not have any  $b-$tagged jet veto which will reduce $t\bar{t}$ significantly\,\cite{Aad:2014wea}.
\begin{figure}[t]
\centering
 \includegraphics[width=0.45\textwidth]{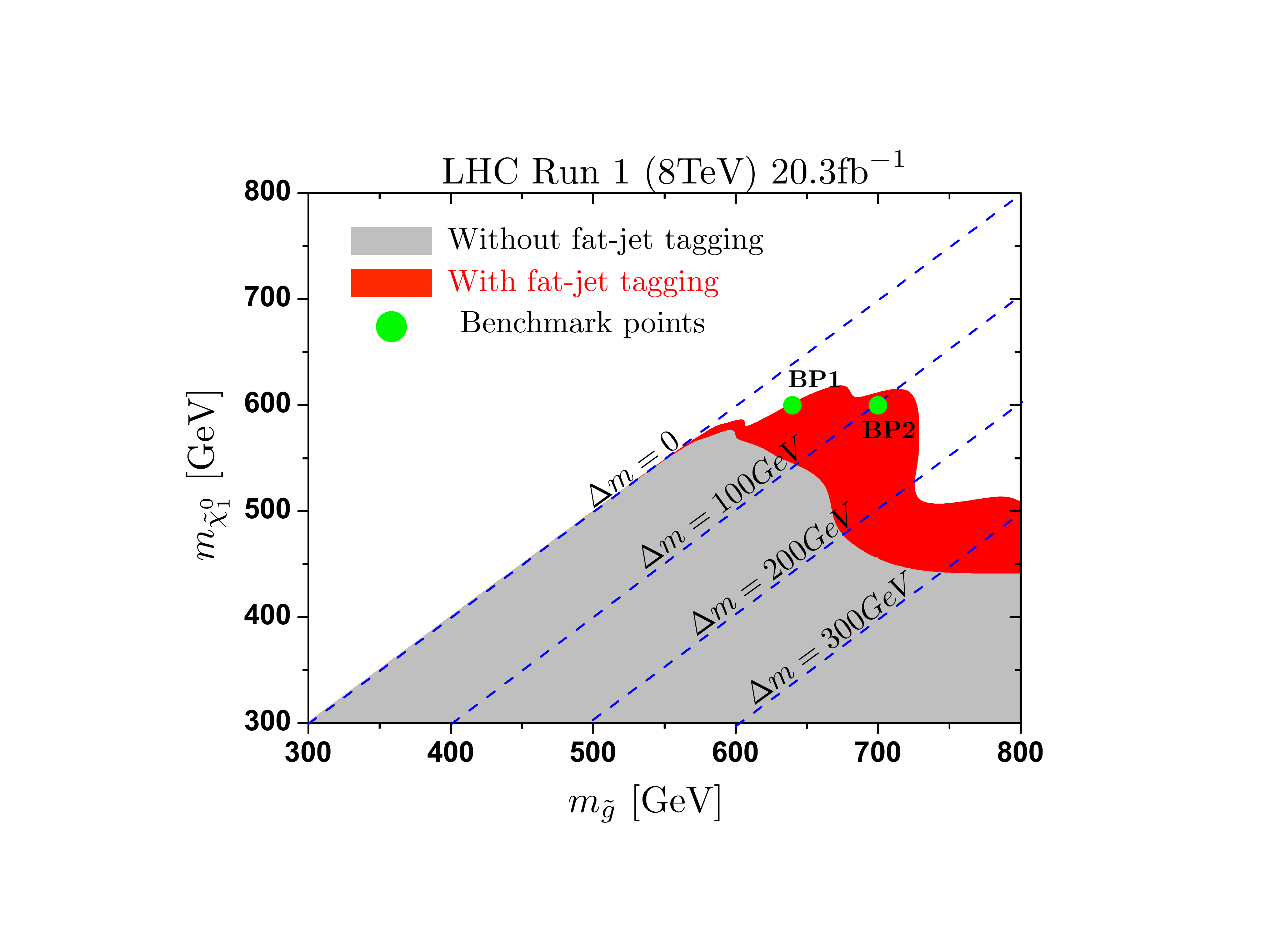}
 \caption{We show an improvement from utilizing fat-jet tagging on search sensitivity of $2\sigma$ in multi-jet signal region. }
\label{fig:exclusion}
\end{figure}

To see overall enhancement of the search sensitivity, we perform a complete scanning over $(m_{\tilde g},\, m_{\tilde \chi_1^0})$ as in FIG.\,(\ref{fig:exclusion}). 
For a fat-jet analysis, we impose $\rho > 0.1$ cut and fat-jet mass window cut of
\bea
\frac{1}{4} \Delta m < m_{\textrm{FJ}} < \Delta m \textrm{ for } \Delta m \ge 20\GeV\, , \\
5\GeV <  m_{\textrm{FJ}} < 20\GeV \textrm{ for } \Delta m < 20\GeV\, .
\eea
With the limited luminosity of the LHC Run 1, we gain only small increasement in the very degenerated case of $\Delta m\lesssim20\GeV$. This limit is from the jet qualification of $p_T > 20\GeV$ in pile-up removal \cite{CMS:2014ata}.
With upcoming LHC runs, we will access the very degenerated region with a moderate help from ISR jet to have more fat-jet above the jet qualification.
For the $14\TeV$ LHC with a luminosity $30\textrm{fb}^{-1}$, we confirm the enhancement from fat-jet preselection as a significance of $2.4$ compared to the result of conventional ATLAS analysis $\sigma\sim1.64$ in the studypoint of $\Delta m = 20\GeV$ with $m_{\tilde \chi_1^0}=700\GeV$. 
In analyses using large size jets, pile-up may give effects on the filtered jet by smearing jet mass. But we note that pile-up would not affect our results significantly based on $7\TeV$ ATLAS analysis where the rate on the change of large size jet mass after MDT,
$\frac{dm}{dN_{PV}}=0.1\pm0.2\GeV$ up to $\mathcal{O}(10)$ pile-ups\,\cite{jet-atlas}.

{\it Conclusion.}---We have proposed a method of clustering soft jets from decaying process with a fat-jet in the scenario of the compressed spectrum.
To identify signals over backgrounds, we apply a jet grooming method and a jet substructure variable to fat-jets. 
 Tagging particles in signals will enable us to probe the properties (mass, spin and coupling structure) with upcoming high luminosity LHC\,\cite{Lester:1999tx, Cho:2007qv, Mahbubani:2012kx, Edelhauser:2012xb}. There have been studies to use a tagged jet through jet substructure methods 
 to see the possibility whether tagged jet maintains the parton level information \cite{parton}. 
 More detailed study in this direction should be performed once the LHC observes the signal. 

\begin{acknowledgments}
The earlier stage of this work was supported by the Korea Ministry of Science, ICT and Future Planning,
Gyeongsangbuk-Do and Pohang City for Independent Junior Research
Groups at the Asia Pacific Center for Theoretical Physics. MP is supported by IBS under the project code, IBS-R018-D1.
CCH appreciates Michihisa Takeuchi's help to generate MC samples using Herwig. MP thanks Steven Mrenna for the information on mcplots.
CCH is supported by World Premier International Research Center Initiative
(WPI Initiative), MEXT, Japan.
\end{acknowledgments}

\end{document}